\documentclass[aps,prl,reprint,superscriptaddress,showpacs]{revtex4-1}
\usepackage{amsmath}
\usepackage{graphicx}% Include figure files
\usepackage{dcolumn}% Align table columns on decimal point
\usepackage{bm}% bold math

\begin{document}

% Use the \preprint command to place your local institutional report
% number in the upper righthand corner of the title page in preprint mode.
% Multiple \preprint commands are allowed.
% Use the 'preprintnumbers' class option to override journal defaults
% to display numbers if necessary
%\preprint{}

%Title of paper
\title{Charge-changing-cross-section measurements of $^{12-16}$C at around $45A$ MeV and development of a Glauber model for incident energies $10A-2100A$ MeV}

% repeat the \author .. \affiliation  etc. as needed
% \email, \thanks, \homepage, \altaffiliation all apply to the current
% author. Explanatory text should go in the []'s, actual e-mail
% address or url should go in the {}'s for \email and \homepage.
% Please use the appropriate macro foreach each type of information

% \affiliation command applies to all authors since the last
% \affiliation command. The \affiliation command should follow the
% other information
% \affiliation can be followed by \email, \homepage, \thanks as well.
\author{D.T.~Tran}
\email[]{tdtrong@rcnp.osaka-u.ac.jp}
%\homepage[]{Your web page}
%\thanks{}
%\altaffiliation{}
\affiliation{Research Center for Nuclear Physics, Osaka University, Osaka, Japan.}
\affiliation{Institute Of Physics, Vietnam Academy of Science and Technology, Hanoi, Vietnam.}

\author{H.J.~Ong}
\email[]{onghjin@rcnp.osaka-u.ac.jp}
\affiliation{Research Center for Nuclear Physics, Osaka University, Osaka, Japan.}

\author{T.T.~Nguyen}
\affiliation{Pham Ngoc Thach University of Medicine, HCM, Vietnam}
\affiliation{Faculty of Physics and Engineering Physics, VNU-HCMUS, HCM, Vietnam.}

\author{I.~Tanihata}
\affiliation{Research Center for Nuclear Physics, Osaka University, Osaka, Japan.}
\affiliation{School of Physics and Nuclear Energy Engineering and IRCNPC, Beihang University, Beijing, China.}

\author{N.~Aoi}
\affiliation{Research Center for Nuclear Physics, Osaka University, Osaka, Japan.}

\author{Y.~Ayyad}
\affiliation{Research Center for Nuclear Physics, Osaka University, Osaka, Japan.}

\author{P.Y.~Chan}
\affiliation{Research Center for Nuclear Physics, Osaka University, Osaka, Japan.}

\author{M.~Fukuda}
\affiliation{Department  of Physics, Osaka University, Osaka, Japan.}

\author{T.~Hashimoto}
\affiliation{Institute for Basic Science, Daejeon, Korea.}
 
\author{T.H.~Hoang}
\affiliation{Research Center for Nuclear Physics, Osaka University, Osaka, Japan.}
\affiliation{Institute Of Physics, Vietnam Academy of Science and Technology, Hanoi, Vietnam.}

\author{E.~Ideguchi}
\affiliation{Research Center for Nuclear Physics, Osaka University, Osaka, Japan.}

\author{A.~Inoue}
\affiliation{Research Center for Nuclear Physics, Osaka University, Osaka, Japan.}

\author{T.~Kawabata}
\affiliation{Department  of Physics, Kyoto University, Kyoto, Japan.} 

\author{L.H.~Khiem}
\affiliation{Institute Of Physics, Vietnam Academy of Science and Technology, Hanoi, Vietnam.}

\author{W.P.~Lin}
\affiliation{Institute of Modern Physics, Lanzhou, China.}

\author{K.~Matsuta}
\affiliation{Department  of Physics, Osaka University, Osaka, Japan.}

\author{M.~Mihara}
\affiliation{Department  of Physics, Osaka University, Osaka, Japan.}

\author{S.~Momota}
\affiliation{Kochi University of Technology,  Kochi, Japan.}

\author{D.~Nagae}
\affiliation{Institute of Physics, University of Tsukuba, Tsukuba, Japan.}

\author{N.D.~Nguyen}
\affiliation{Dong Nai University, Dong Nai, Vietnam.}

\author{D.~Nishimura}
\affiliation{Tokyo University of Science, Tokyo, Japan.}

\author{A.~Ozawa}
\affiliation{Institute of Physics, University of Tsukuba, Tsukuba, Japan.}
 
\author{P.P.~Ren}
\affiliation{Institute of Modern Physics, Lanzhou, China.}

\author{H.~Sakaguchi}
\affiliation{Research Center for Nuclear Physics, Osaka University, Osaka, Japan.}

\author{J.~Tanaka}
\affiliation{Research Center for Nuclear Physics, Osaka University, Osaka, Japan.}

\author{M.~Takechi}
\affiliation{Graduate School of Science and Technology, Niigata University, Niigata, Japan.}

\author{S.~Terashima}
\affiliation{School of Physics and Nuclear Energy Engineering and IRCNPC, Beihang University, Beijing, China.}

\author{R.~Wada}
\affiliation{Cyclotron Institute, Texas A$\&$M University, College Station, Texas, USA.}
\affiliation{Institute of Modern Physics, Lanzhou, China.}

\author{T.~Yamamoto}
\affiliation{Research Center for Nuclear Physics, Osaka University, Osaka, Japan.}

\collaboration{RCNP-E372 collaboration}

\date{\today} 
\begin{abstract}
  % insert abstract here
  We have measured for the first time the charge-changing cross sections ($\sigma_{\text{\tiny CC}}$) of
  $^{12-16}$C on a $^{12}$C target at energies below $100A$ MeV. To analyze these low-energy data,
  we have developed a finite-range Glauber model with a global parameter set within the optical-limit approximation
  which is applicable to reaction cross section ($\sigma_{\text{\tiny R}}$) and $\sigma_{\text{\tiny CC}}$ measurements
  at incident energies from 10$A$ to $2100A$ MeV. Adopting the proton-density distribution of
  $^{12}$C known from the electron-scattering data, as well as the bare total nucleon-nucleon cross
  sections, and the real-to-imaginary-part ratios of the forward proton-proton elastic scattering
  amplitude available in the literatures, we determine the energy-dependent slope parameter
  $\beta_{\rm pn}$ of the proton-neutron elastic differential cross section so as to reproduce the
  existing $\sigma_{\text{\tiny R}}$ and interaction-cross-section data for $^{12}$C+$^{12}$C over a wide
  range of incident energies. The Glauber model thus formulated is applied to calculate the
  $\sigma_{\text{\tiny R}}$'s of $^{12}$C on a $^9$Be and $^{27}$Al targets at various incident energies. Our 
  calculations show excellent agreement with the experimental data. Applying our
  model to the $\sigma_{\text{\tiny R}}$ and $\sigma_{\text{\tiny CC}}$ for the ``neutron-skin'' $^{16}$C nucleus,
  we reconfirm the importance of measurements at incident energies below $100A$ MeV. The proton
  root-mean-square radii of $^{12-16}$C are extracted using the measured $\sigma_{\text{\tiny CC}}$'s and
  the existing $\sigma_{\text{\tiny R}}$ data. The results for $^{12-14}$C are consistent with the
  values from the electron scatterings, demonstrating the feasibility, usefulness of the
  $\sigma_{\text{\tiny CC}}$ measurement and  the present Glauber model.
\end{abstract}

% insert suggested PACS numbers in braces on next line
% insert suggested keywords - APS authors don't need to do this
%\keywords{}

%\maketitle must follow title, authors, abstract, \pacs, and \keywords
\maketitle

% body of paper here - Use proper section commands
% References should be done using the\cite, \ref, and \label commands
\section{Introduction}
The nuclear sizes, usually defined by the root-mean-square (rms) charge or nucleon/matter distribution 
radii, are important nuclear quantities. The proton and neutron rms radii are not only important to 
extract information on the nuclear structure, but are also essential for extracting the neutron skin 
thickness, which offers an important means to constrain theoretical descriptions of the equation of 
state (EOS) of asymmetric nuclear matter~\cite{BrownPRL85}. The nuclear EOS is important to understand 
the properties of dense nuclear matter such as the neutron stars as well as to predict supernovae and 
neutron star mergers~\cite{Lattimer62}. \\
\indent Historically, the earliest evidence for a nuclear radius came not from a direct measurement,
but was inferred from the studies of the $\alpha$ decay of radioactive nuclei~\cite{Gamow}. It was 
only after 1950s, with the advent of particle accelerators and the quantum electrodynamic theory, 
that decisive evidences for finite nuclear sizes and more precise measurements of charge/proton
radii became available. Scores of charge radii of mostly stable nuclei have since been precisely 
determined using electromagnetic probes such as the elastic scattering of fast electrons, 
X-ray spectroscopy of muonic atom, optical and K$_\alpha$ X-ray isotope-shift (IS) methods~\cite{Elton}.\\
\indent For short-lived unstable nuclei, the IS method had been the only source of information
until very recently. The electron scattering which has been the most successful method to determine
the nuclear charge radii is not applicable because the short-lived nuclei are not available as targets.
While the effort to perform electron scattering on unstable nuclei is being 
pursued~\cite{WakasugiNIMB317}, it may take some time to achieve practical applications.
The optical IS method, on the other hand, requires only a small number of atoms of
the unstable nuclei. Experimentally, the IS measurements using laser spectroscopy have achieved 
very high precision (below 100 kHz) and sensitivity~\cite{CampbellPPNP86}. Spurred on by recent 
advances in computational methods, the IS methods have been successfully applied to determine 
the charge radii of light unstable nuclei up to 
$^{12}$Be~\cite{WangPRL93,MuellerPRL99,SanchezPRL96,NorterPRL102,TakaminePRL112,KriegerPRL108}.
However, it is extremely challenging to apply the IS method to the $10>Z>4$ nuclei due mainly to 
insufficient precision in the atomic physics calculations and difficulty of production of low-energy isotopes. \\
\indent In terms of other non-electromagnetic probe, an important breakthrough was achieved in
1985 through the measurements of interaction cross sections of light neutron-rich nuclei, which led
not only to the discovery of the neutron-halo structure~\cite{TanihataPRL55} but also to the
renaissance in nuclear physics with radioactive beams.  Applying the Glauber model~\cite{Glauber},
the nuclear matter rms radii of neutron-rich He, Li and Be isotopes were extracted for the first
time~\cite{TanihataPRL55}. Since then, interaction ($\sigma_{\text{\tiny I}}$) as well as reaction cross
sections ($\sigma_{\text{\tiny R}}$) have been extensively measured, providing a wealth of information on
the rms radii of the nuclear matter distribution of unstable nuclei up to the proton and neutron
driplines~\cite{OzawaNPA693}. Recently, extending the Glauber-type analysis to the measured
charge-changing cross section ($\sigma_{\text{\tiny CC}}$), which is the total cross sections of all processes
that change the proton number of a nucleus, I.~Tanihata demonstrates~\cite{TanihataPPNP68}, through comparisons with
the results from the IS method, the feasibility of the $\sigma_{\text{\tiny CC}}$ measurements to determine
the point-proton distribution rms radii (referred to as ``proton rms radii''
hereinafter). Combining $\sigma_{\text{\tiny R}}$ and $\sigma_{\text{\tiny CC}}$ (or the proton rms radii determined by
other electromagnetic probe), it is possible to determine the neutron distribution rms radii.
The successful applications of the method to neutron-rich Be~\cite{TerashimaPTEP101},
B~\cite{EstradePRL113} and C~\cite{YamaguchiPRT107} isotopes at incident-beam energy higher than
$200A$ MeV mark an important milestone in the studies of nuclear radii. \\
\indent The Glauber model has been the most widely used and successful method to determine matter
rms radii of unstable nuclei. However, the applicability of this method at low-incident energies
has been questionable. While the optical-limit approximation (OLA) of the Glauber model under
the zero-range approximation (ZR) has proven to be the most economic and convenient model to
calculate $\sigma_{\text{\tiny I}}$ or $\sigma_{\text{\tiny R}}$ at high incident energies~\cite{OzawaNPA693},
it failed to reproduce the experimental data at energies below $100A$ MeV. The discrepancy reaches
almost 20$\%$ at a few tens of MeV per nucleon for the carbon isotopes~\cite{TakechiEPJA25}. This
discrepancies could be due to various possible effects such as the Fermi motion, Pauli
correlations~\cite{DiGiacomo}, and short-range dynamic correlations. Taking into account the
higher-order multiple scattering and Fermi-motion effects, M.~Takechi {\it et al.}~\cite{TakechiPRC79} modified the
bare nucleon-nucleon interaction cross sections and obtained calculations that reproduce the
experimental $\sigma_{\text{\tiny R}}$'s relatively well over a wide range of incident
energies. B.~Abu-Ibrahim and Y.~Suzuki~\cite{IbrahimPRC62}, on the other hand, pointed out that the
above-mentioned various effects would have been automatically included to some extent in formulating
the profile function for the {\it N--N} scatterings.\\
\indent In this paper, we report on the first measurement of the Charge-Changing cross sections
($\sigma_{\text{\tiny CC}}$) of $^{12-16}$C on a $^{12}$C target at incident energies at around $45A$ MeV.
To analyze the data and extract the proton rms radii, we have developed a Glauber model within
the optical-limit approximation (OLA), which is applicable to a wide energy range between $10A$
and $2100A$ MeV. Here, we determine the energy-dependent slope parameter $\beta_{\rm pn}$ of the proton-neutron
elastic differential cross section, which is the only missing parameter besides the
density distributions required in the Glauber model calculation. The $\beta_{\rm pp}$ parameter values for
 proton-proton scattering were adopted from the proton-proton scattering data.
 The extension of Glauber model to energies below $100A$ MeV is important because of
the sensitivity of the low-energy $\sigma_{\text{\tiny R}}$ (and perhaps $\sigma_{\text{\tiny CC}}$) to the tail-density
distributions of halo and skin nuclei. Such sensitivity has been demonstrated by the
$\sigma_{\text{\tiny R}}$'s of $^{11}$Be on a $^{12}$C and a $^{27}$Al~\cite{FukudaPLB268}, as well as of
$^{16}$C on a $^{12}$C target~\cite{ZhengNPA709}. Applying the present Glauber model to calculate the
reaction cross sections of the $^{12}$C on a $^9$Be and $^{27}$Al targets, we demonstrate the
reliability of our model. We also show that the extracted proton rms radii for $^{12-14}$C are
consistent with the results from the electron scatterings.

\section{Experiment}
\begin{figure}[tb]
\includegraphics[scale=0.7]{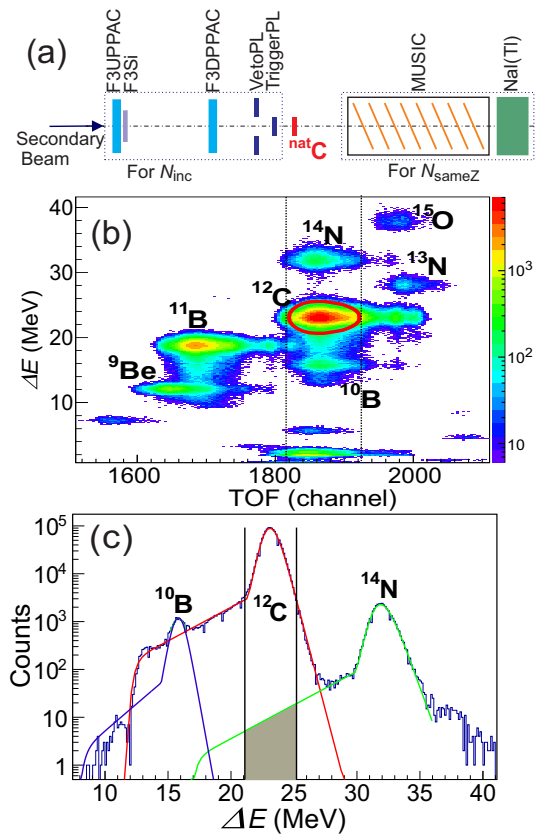}% Here is how to import EPS art
\caption{\label{fig:setup} (Color online). (a) Schematic view of the experiment
setup, (b) in-coming $^{12}$C beam identification, and (c) contaminant estimation.}
\end{figure}
The experiment was performed at the EN course~\cite{ShimodaNIMB70}, Research Center for
Nuclear Physics (RCNP), Osaka University. Secondary $^{12-16}$C beams were produced in separate runs 
by fragmentation of a $^{22}$Ne primary beam at $80A$ MeV incident on a $^9$Be 
target with thickness ranging from 1.0 - 5.0 mm. The carbon isotope of interest was selected 
in flight by setting the appropriate magnetic rigidities of two dipole magnets
of the EN fragment separator. A flat aluminum degrader, with thickness ranging from 0.3 -
5.0 mm, was placed at the first momentum-dispersive focal plane (F1) to improve the isotope
separation of the secondary beams. The momentum acceptances of the secondary beams were typically
set to $\pm 0.2\%$ using a set of collimators at F1. The secondary beams were angular 
focused at the second focal plane (F2), which is a momentum-achromatic and a 
charge-mass dispersive focal plane. The selected carbon-isotope beam was further purified using 
a set of collimators at F2 before being transported to and directed onto a $450$-mg/cm$^2$-thick 
natural carbon target (reaction target) placed at the newly-constructed third focal plane 
(F3)~\cite{OngRCNPAR2011}.

In the present work, we measured the $\sigma_{\text{\tiny CC}}$'s of carbon isotopes employing the 
transmission method. Figure~\ref{fig:setup}(a) shows the experimental setup at F3. The incoming 
carbon-isotope beam was identified on an event-by-event basis using the energy-loss 
($\Delta E$) and time-of-flight (TOF) method. $\Delta E$ was measured using a 320-$\mu$m-thick 
silicon detector, while the TOF between the $^9$Be production target and the reaction target
was determined using the timing information from a 100-$\mu$m-thick plastic scintillator
placed right before the reaction target and the RF signal from the cyclotron. The timing 
signal from the plastic scintillator was also used as the trigger for the data-acquisition 
system. Incident particles were tracked using the position information obtained with four Parallel 
Plate Avalanche Counters (PPACs)~\cite{KumagaiNIMA470} located before F2 and F3. To select and
define ``good'' incident carbon-isotope, we rejected the particles that scattered at 
large angles after the last PPAC using a 3-mm-thick ``veto'' plastic scintillator, which has a 
square hole of a size smaller than the reaction target at its center, placed before the trigger 
scintillator. The number of ``good'' incident particles thus counted is denoted by 
$N_{\text{inc}}$.\\
\indent The outgoing particles went through a MUlti Sampling Ionization Chamber 
(MUSIC)~\cite{KimuraNIMA538}, which consists of eight anodes and nine cathodes, before being 
stopped in a 7-cm-thick NaI(Tl) scintillator. The $\Delta E-E$ method was employed to identify and 
count the scattered particles. The $Z$-unchanged  particles  are counted and denoted by $N_{\text{same}Z}$.
\section{Data Analysis and Results}
In the transmission method, the $\sigma_{\text{\tiny CC}}$ is calculated as follows:
$\sigma_{\text{\tiny CC}}=\text{ln}\left[ \gamma_{\text{\tiny 0}}/\gamma \right]/t$, 
where, $t$ is the number of target nuclei per cm$^2$ of beam area, and $\gamma$ and $\gamma_{\text{\tiny 0}}$ are the ratios of the number of the  $Z$-unchanged particles and the number of incident particles, $\gamma = N_{\text{same}Z}/N_{\text{inc}}$, of measurements with and without the reaction target respectively. 

We determined the $N_{\text{inc}}$ and $N_{\text{same}Z}$ using the information from the detectors
before and after the reaction target respectively. Figure~\ref{fig:setup}(b) shows a typical
$\Delta E$-TOF scatter plot for the secondary beams; the red ellipse shows the
Particle-IDentification (PID) gate for $^{12}$C. The contaminant in the PID gate was mainly the
heavier isotopes with reduced energy losses due to the channelling effect in the silicon crystal.
To estimate the amount of contaminant, we selected the TOF region as shown by the dotted lines in
Fig.~\ref{fig:setup}(b) and projected onto the $\Delta E$ axis.
Fig.~\ref{fig:setup}(c) shows the projected $\Delta E$ distribution for the three nuclides
with long tails due to the channeling effect.  By scaling the distribution in Fig.~\ref{fig:setup}(c) to 
$N_{\text{inc}}$, the contaminant was estimated to be less than 0.6$\%$ of $N_{\text{inc}}$. 
The admixtures were further identified and confirmed using the detectors after the reaction target.
 Depending on the statistics of the carbon isotopes, the contaminants contribute to systematic uncertainties 
of only about $0.1-3.5$ mb in the final cross sections, and are much smaller than the errors of the cross sections. \\
\begin{figure}[bp]
  \includegraphics[scale=0.85]{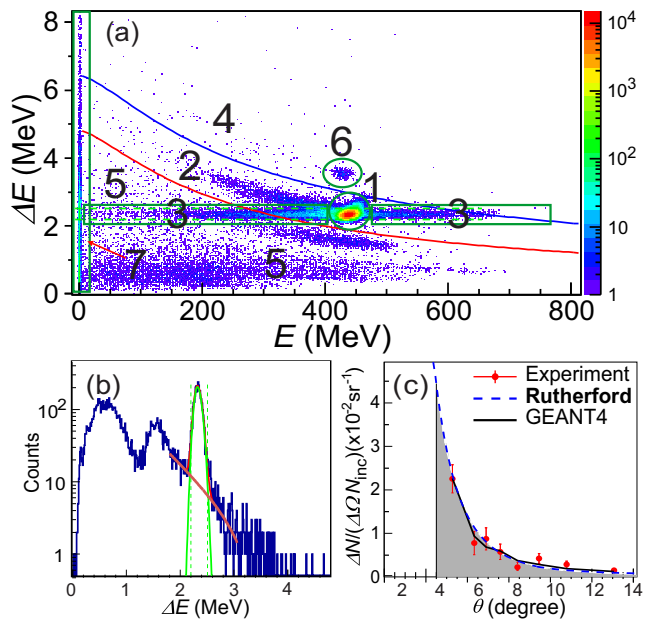}% Here is how to import EPS art
  \caption{\label{fig:pid} (Color online). (a) Identification of the scattered particles.
    (b) Estimation of background contribution from region 5. (c) Acceptance-correction
    factor calculation. For details see text.}
\end{figure}
\indent The detection and particles identification of the $Z$-unchanged  particles in the present 
reaction energies are more complicated than in the high energy due to energy loss 
straggling and multiple scatterings of the outgoing charged particles in the target and detector 
materials. The former results in broadening of the measured energy losses while the latter in 
reduced geometrical acceptance for the scattered-particle detectors. Figure~\ref{fig:pid}(a) shows 
a typical $\Delta E-E$ plot for scattered particles obtained with the MUSIC ($\Delta E$) and 
NaI(Tl) scintillator ($E$). The particles are classified by 7 regions as shown in the figure: 
(1) beam-like particles, (2) elastic and inelastically scattered beam-like particles, (3) particles
that reacted in the NaI(Tl) scintillator, (4) proton-picked-up particles, (5) proton-removed 
particles, (6) beam contaminants, and (7) ``out-of-acceptance'' particles, which were not detected 
by the NaI(Tl) detector. The number of particles with the same $Z$ as the selected incident beam 
was determined by summing the events in the regions 1, 2 and 3. To estimate the number of
light particles in region 3, a Gaussian peak plus an exponential background function was used to 
fit the experimental data (see Fig.~\ref{fig:pid}(b)). The systematic uncertainties
attributed to the background that contribute to the final $\sigma_{\text{\tiny CC}}$'s are bellow 1 mb for
all carbon isotopes. \\
\indent The main source of systematic uncertainties lies in the estimation of the out-of-acceptance
carbon isotopes in the region 7. The particles in the region 7 comprised about $2\%$ of the total 
events. Simply adopting this value as the systematic uncertainty results in as large as $20\%$ 
uncertainty in the measured $\sigma_{\text{\tiny CC}}$. Hence, to reduce this uncertainty, we introduced 
an acceptance-correction factor, denoted by $P$. The final $\sigma_{\text{\tiny CC}}$ was deduced as 
follows:
\begin{equation}
  \sigma_{\text{\tiny CC}}=\frac{1}{t} ln\left[\frac{\gamma_{\text{out}}(1-P_{\text{in}})}{\gamma_{\text{in}}(1-P_{\text{out}})}\right]
  \label{qe:cccs}
\end{equation}
where the subscripts ``in'' and ``out'' indicate measurements with and without reaction target, 
respectively. $P_{\text i}$ (i = in or out) was determined by assuming the scattering at large angle
as being mainly due to the Rutherford scattering. To determine the experimental $P_{\text i}$, we
first calculated the difference in the solid angles ($\Delta \Omega$) covered by a particular MUSIC
electrode and the next layer (a MUSIC electrode or the NaI(Tl) scintillator) using the geometrical 
information of the experimental setup. By taking the event having an appropriate signal in one layer 
of the MUSIC but not in the next layer as the event being scattered into the solid angle 
$\Delta \Omega$, the number of lost events $\Delta N$ was determined for each scattering angle. 
The $\Delta N/(\Delta \Omega N_{\text{inc}})$ ratios thus obtained are proportional to the 
differential cross sections of the elastic Coulomb scattering, and were fitted with a calculated 
Rutherford scattering differential cross section distribution. As shown in the 
Fig.~\ref{fig:pid}(c), the experimental data are well reproduced by the Rutherford distribution. To
further confirm the assumption, we performed Monte Carlo simulations using the Geant4
code~\cite{AgostinelliNIMA506}. The results from the simulations are also in excellent agreement
with the experimental data as well as the Rutherford distribution. The $P_{i}$ value is simply the
integral of the distribution over the solid angles not covered by the NaI(Tl) detectors, as shown
by the shaded area in Fig.~\ref{fig:pid}(c). Depending on isotope, the $P_{\text{in}}$
($P_{\text{out}}$) value thus determined varies from 0.003 (0.0005) to 0.004 (0.0012), with an
uncertainty between 2 -- 10$\%$ (5 -- 15$\%$). This uncertainties contribute to 6 -- 10 mb of
$\sigma_{\text{\tiny CC}}$ for different isotopes.\\
\indent The determined $\sigma_{\text{\tiny CC}}$ values are summarized in Table~\ref{tab:cccs}. The 
uncertainties (in brackets) include the above-mentioned systematic uncertainties, the statistical 
uncertainties as well as the uncertainty in the target thickness ($0.06\%$). The results for 
$^{12}$C at $38.0A$-MeV incident energy, measured during the same experiment to examine possible 
systematic uncertainty due to the incident-beam energy, are also shown. 
\begin{table}[t]
\caption{\label{tab:cccs}
  The experimental values of $\sigma_{\text{\tiny CC}}$ and proton, neutron rms radii of carbon
  isotopes.}
\begin{ruledtabular}
\begin{tabular}{cdrdd}
 &\multicolumn{1}{c}{\textrm{E(MeV/u)}}
 &\multicolumn{1}{c}{\textrm{$\sigma_{\text{\tiny CC}}$ (mb)}}
 &\multicolumn{1}{r}{\textrm{r$_{\text{p}}^{\text{exp}}$(fm)}}
 &\multicolumn{1}{r}{\textrm{r$_{\text{p}}^{\text{ref}}$(fm)}}\\
 \hline
$^{12}$C& 38.0 & 1056(20)\\
$^{12}$C& 48.4 & 941(16) & 2.35(6)\footnotemark[1] &2.327(7)\footnotemark[2] \\ 
$^{13}$C& 47.7 & 968(39) & 2.35(9) &2.321(8)\footnotemark[2] \\ 
$^{14}$C& 46.3 & 960(18) & 2.32(4) & 2.370(11)\footnotemark[2] \\ 
$^{15}$C& 44.1 & 987(34) & 2.41(8)&2.33(11)\footnotemark[3] \\ 
$^{16}$C& 44.9 & 987(20) & 2.40(5)&2.25(11)\footnotemark[3] \\ 
\end{tabular}
\end{ruledtabular}
\footnotetext[1]{The average value of two energies is shown.}
\footnotetext[2]{From Ref.~\cite{AngeliADNDT99}.}
\footnotetext[3]{From Ref.~\cite{YamaguchiPRT107}.}
\end{table}
\section{Formulation of the Glauber Model}
To extract the proton rms radii, we performed finite-range Glauber-model calculations within the 
OLA using the parameter set from nucleon--nucleon ({\it N--N}) cross sections.
 Following the procedures in Ref.~\cite{TerashimaPTEP101} and 
ignoring the effect of neutrons in a projectile, we calculate $\sigma_{\text{\tiny CC}}$ as follows: 
\begin{equation}
\sigma_{\text{\tiny CC}}=2\pi\int{d\vec{b} \left[1-|{\text e}^{{\text i}\chi({b})}|^2\right]}
\label{qe:glauber}
\end{equation}
where $\vec{b}$ is the impact parameter, and the exponential term is the transmission function 
given by the following relation:
%$$
\begin{equation*}
\begin{split}
&{\text e}^{{\text i}\chi({b})}= \\
 & {\text{exp}}\left[\int_{\rm P} \int_{\rm T} {\sum\limits_{N} {\left[ {\rho _{{\text{\tiny P}}_{\rm p}}^z\left( \vec{ s} 
\right) \rho _{{\text{\tiny T}}_{\text{\tiny{\it N}}}}^z\left( \vec{t} \right)\Gamma_{{\rm p}{\text{\tiny{\it N}}}}\left( {\vec{b} + \vec{s} - \vec{t}} 
\right)} \right]d\vec{s}d\vec{t}} }\right].
\end{split}
\end{equation*}
%$$
The superscript $z$ in the above formula indicates the direction of integration, which corresponds to the direction of the incident particle, 
for the nucleon density. $\rho _{{\text{\tiny P}}_{\rm p}}^z$ is the proton density of the 
projectile and $\rho _{{\text{\tiny T}}_N}^z$ with subscript $N={\text{p,n}}$ is the proton or neutron
 density of the target. $\vec {s}$ (${\vec t}$) represents the two-dimensional 
coordinate of a particular projectile (target) nucleon relative to the center of mass of the 
projectile (target) nucleus, which lies on the plane perpendicular to the incident momentum of the 
projectile. $\Gamma$ is the {\it N--N} amplitude~\cite{RayPRC20}, which in the case of 
the scatterings of protons off a nuclear target simplifies as the profile 
function~\cite{IbrahimPRC62}:
\begin{equation}
 \Gamma_{{\rm p}{\text{\tiny{\it N}}}}\left( {\vec b} \right) = \frac{1-i\alpha_{\text{p{\tiny{\it N}}}}}{{4\pi \beta _{\text{p{\tiny{\it N}}}}}}
\sigma^{\rm tot}_{\text{p{\tiny{\it N}}}}{\text{exp}}\left[-{\frac{{{b^2}}}{{2\beta _{\text{p{\tiny{\it N}}}}}}} 
\right]
\end{equation}
where $\alpha_{\text{p{\tiny{\it N}}}}$ is the ratio of the real to the imaginary part of the forward p--{\it N} 
scattering amplitude, $\beta_{\text{p{\tiny{\it N}}}}$ the slope parameter of the p--{\it N} elastic differential 
cross section, and $\sigma^{\rm tot}_{\text{p{\tiny{\it N}}}}(E)$ is the total p--{\it N} cross section at incident 
energy $E$. The energy-dependent $\alpha_{\text{p{\tiny{\it N}}}}$ and $\beta_{\text{p{\tiny{\it N}}}}$ parameters are 
interrelated, via the total elastic cross section ($\sigma_{\text{p{\tiny{\it N}}}}^{\text{el}}(E)$) and 
$\sigma^{\rm tot}_{\text{p{\tiny{\it N}}}}(E)$, as follows~\cite{OgawaNPA543}:
\begin{equation}
  \sigma_{\text{p{\tiny{\it N}}}}^{\text{el}}(E)=\frac{1+\alpha_{\text{p{\tiny{\it N}}}}^2}{16\pi\beta_{\text{p{\tiny{\it N}}}}}\left[\sigma^{\rm tot}_{\text{p{\tiny{\it N}}}}(E)\right]^2.
\label{qe:alphabeta}
\end{equation}

\begin{figure}[bp]
\includegraphics[scale=0.5]{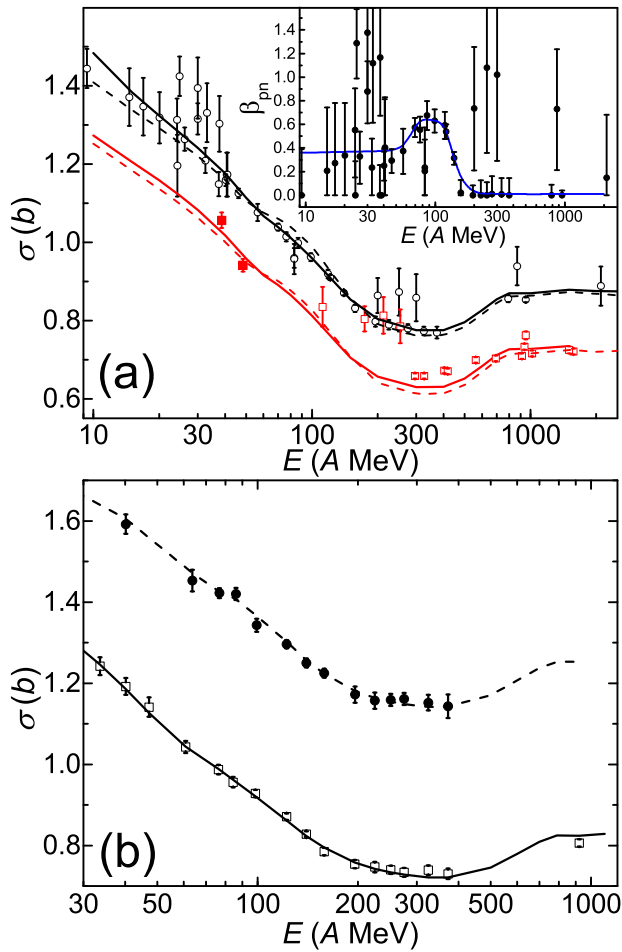}% Here is how to import EPS art
\caption{\label{fig:glauber} (Color online). (a) Experimental $\sigma_{\text{\tiny R}}$'s (black-open 
  circles)~\cite{TakechiPHD,OzawaNPA691} and $\sigma_{\text{\tiny CC}}$'s (red symbols) of $^{12}$C on carbon target. The red-filled 
  and red-open squares are our data at $45A$ MeV and data from Ref.~\cite{WebberPRC41,ChulkovNPA674}
  respectively. The black solid line is the energy-dependent $\sigma_{\text{\tiny R}}$($E$) calculated 
  with the best-fitted $\beta_{\rm pn}$($E$) (inset) and HO-type neutron density distribution. 
  The red solid line is the calculated $\sigma_{\text{\tiny CC}}$($E$). The dashed (black and red) lines 
  are results of Glauber calculations (for $\sigma_{\text{\tiny R}}$($E$) and $\sigma_{\text{\tiny CC}}$($E$))
  with the OLA plus higher-order correction~\cite{IbrahimPRC61}. (b) Experimental 
  $\sigma_{\text{\tiny R}}$'s for $^{12}$C on a beryllium (open squares) and aluminum (filled circles) 
  targets. The solid and dashed lines are the $\sigma_{\text{\tiny R}}$($E$)'s calculated with the present
    Glauber model.  A neutron-density distribution with a tail structure is necessary for beryllium target. 
    See text for the details on the input density distributions for the target 
  nucleons.}
\end{figure}

\indent In the OLA calculation, only the real part of the profile function that contains only the 
$\beta_{\text{p{\tiny{\it N}}}}$ parameter contributes to the cross section. Hence, it is sufficient to determine 
$\beta_{\text{pp}}$ and $\beta_{\text{pn}}$  for the Glauber model calculations. Substituting the $\alpha_{\text{pp}}$ values and the cross 
sections from the Particle Data Group tabulation~\cite{OliveCPC38} into Eq.~\ref{qe:alphabeta}, we 
deduced $\beta_{\text{pp}}$ over a wide range of incident energy. For $\beta_{\text{pn}}$, only a few 
data points for $\alpha_{\text{pn}}$ at incident energies above $174A$ MeV are available from 
Ref.~\cite{OliveCPC38}. Although parameter sets from the studies on proton-nucleus scatterings at 
proton energies ranging from 100 to 2200 MeV~\cite{RayPRC20}, and on heavy-ion scatterings at 
projectile energies 30$A$ -- 350$A$ MeV~\cite{LenziPRC40} are available, both parameter sets failed 
to reproduce the energy dependence of the reaction/interaction cross section of 
$^{12}$C~\cite{TakechiPRC79,HoriuchiPRC75}. Introducing separate parametrization schemes for 
energies below and above $300A$ MeV, and adopting partially or modifying the parameters in 
Ref.~\cite{RayPRC20}, several authors have reported improved global 
systematics~\cite{TakechiPRC79,IbrahimNPA657,IbrahimPRC77,HoriuchiPRC75}. \\
\indent In this work, we took a different approach and determined the energy-dependent 
$\beta_{\text{pn}}$($E$), taking advantage of the accumulating experimental 
$\sigma_{\text{\tiny R}}$'s~\cite{TakechiPRC79} of $^{12}$C on a $^{12}$C target at incident energies from 
$10A$ MeV up to about $2100A$ MeV. To this end, we first fixed the proton- and neutron-density 
distributions which are needed for the OLA Glauber calculations. We adopted the sum-of-Gaussian 
distribution from the electron scattering data~\cite{DeVriesADNDT36} as the proton density
distribution in the $^{12}$C target. For the neutrons, assuming a harmonic-oscillator (HO) type 
density distribution, we determined the HO width parameter together with $\beta_{\text{pn}}$ so as 
to reproduce the experimental $\sigma_{\text{\tiny R}}$~\cite{OzawaNPA691} and
$\sigma_{\text{\tiny CC}}$~\cite{WebberPRC41} of $^{12}$C on a carbon target at around $950A$ MeV.
We chose the data at this energy since the Glauber model is well established for high energies.
Using these proton- and neutron-density distributions, we determined the $\beta _{\text{pn}}(E)$ so
as to reproduce the experimental $\sigma_{\text{\tiny R}}$ at various incident energies. The experimental
$\sigma_{\text{\tiny R}}$ data (black-open circles) and the ``fitted'' Glauber model calculation results
(black solid line) are shown in Fig.~\ref{fig:glauber}(a). The best-fitted $\beta_{\rm pn}$($E$) is
shown in the inset.
%
% Figure 3
%
\section{Results of the Glauber-Model Analysis and Discussion}
We applied the Glauber model to calculate the $\sigma_{\text{\tiny CC}}$'s at other energies. 
As shown in Fig.~\ref{fig:glauber}(a), the results show good agreement with the experimental 
$\sigma_{\text{\tiny CC}}$'s in the whole energy range including our measurements (red-filled squares). 
Using the same $\beta _{\text{pn}}(E)$ and density distribution of $^{12}$C, we also calculated $\sigma_{\text{\tiny R}}(E)$ for $^{12}$C on 
beryllium and aluminum targets. Again, we adopted the shape of distribution suggested from the 
electron-scattering data~\cite{DeVriesADNDT36} for the proton density distributions of $^{9}$Be and $^{27}$Al. For the 
neutrons, we assumed a harmonic-oscillator (HO) plus Woods-Saxon (WS) shape for the Be  and a WS shape density 
distributions for the Al target nuclei, namely:
\begin{equation}
\rho_{\text{\tiny Be}} = \rho_{\text{\tiny HO}}(N{\rm =4},R_{\text{\tiny Be}},r) + \rho_{\text{\tiny WS}}(N{\rm =1},R_{\text{\tiny Be}},a_{\text{\tiny Be}},r),
\end{equation}
\begin{equation}
\rho_{\text{\tiny Al} }= \rho_{\text{\tiny WS}}(N{\rm =14},R_{\text{\tiny Al}},a_{\text{\tiny Al}},r),
\end{equation}
where 
\begin{equation*}
\rho_{\text{\tiny HO}}(N,R,r) = \rho_\text{\tiny 0}^{\text{\tiny HO}} \exp \left[ -\left(\frac{r}{R} \right)^2 \right] \left[ 1 + \frac{N-2}{3}\left(\frac{r}{R} \right)^2 \right],
\end{equation*}
\begin{equation*}
\rho_{\text{\tiny WS}}(N,R,a,r) = \frac{\rho_\text{\tiny 0}^{\text{\tiny WS}}}{1 + \exp \left[ (r-R)/a \right]}
\end{equation*}
$\rho_\text{\tiny 0}^{\text{\tiny HO}}$ and $\rho_\text{\tiny 0}^{\text{\tiny WS}}$ are normalization factors that conserve number of neutron(s).
Here, we introduced the Woods-Saxon distribution with a tail density to account for the 
loosely-bound valence neutron in $^9$Be, and determined the diffuseness as well as the HO width 
parameter for the $^9$Be target so as to reproduce the experimental data at 
33.6$A$~\cite{TakechiPHD} and 921$A$ MeV~\cite{OzawaNPA691}. For the $^{27}$Al target, 
$\sigma_{\text{\tiny R}}$'s at 40.2$A$ and 372.4$A$ MeV~\cite{TakechiPHD} were used to determine the WS 
parameters. As shown in Fig.~\ref{fig:glauber}(b), our calculations are in excellent agreement with 
the experimental data of $^{12}$C on beryllium and aluminum targets. We note that calculations 
using the formulation~\cite{IbrahimPRC61} that includes higher order corrections to the OLA yield 
only slightly different results which are consistent with the OLA calculations within the 
experimental uncertainties (see the dashed lines in Fig.~\ref{fig:glauber}(a)).\\

\begin{figure}[bp]
\includegraphics[scale=0.53]{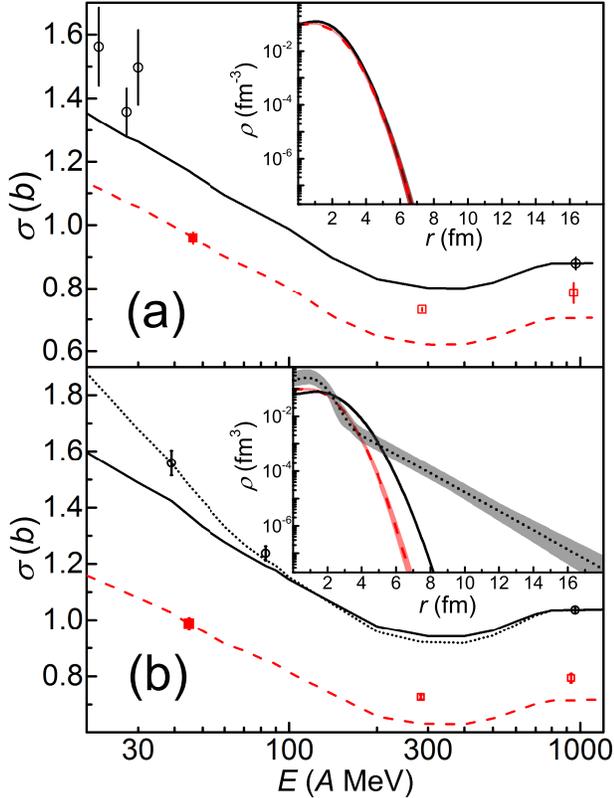}% Here is how to import EPS art
\caption{\label{fig:cccsc} (Color online). Experimental $\sigma_{\text{\tiny CC}}$'s (red symbols) and 
  $\sigma_{\text{\tiny R}}$'s (black symbols) of (a) $^{14}$C and (b) $^{16}$C on a $^{12}$C target. 
  The red-filled squares are our data at around $45A$ MeV. The red-open squares are the data 
  taken from Refs.~\cite{YamaguchiPRT107,ChulkovNPA674}. The black-open circles are the
  $\sigma_{\text{\tiny R}}$ data taken from Refs.~\cite{TakechiPHD,OzawaNPA691}. The red-dashed and the
  black lines correspond to the $\sigma_{\text{\tiny CC}}$($E$) and $\sigma_{\text{\tiny R}}$($E$) calculated using the
  present Glauber model. We assumed HO-type density distributions for the protons and neutrons
  in $^{14}$C, as well as for the protons in $^{16}$C. For $^{16}$C, the neutron-density
  distribution with a HO core plus WS-type-two-neutron tail (core+2n) is necessary to reproduce
  the $\sigma_{\text{\tiny R}}$ at energy below $100A$ MeV. The proton- and neutron-density
  distributions thus determined are shown in the insets: the red-dashed (black-solid) lines
  represent the HO-type proton- (neutron-) density distributions, and the black-dotted line
  represent the core+2n neutron-density distribution. The hatched areas represent the uncertainties in the density distributions due to the uncertainties in the measured cross sections.}
\end{figure}

\indent Figure~\ref{fig:cccsc} shows the experimental $\sigma_{\text{\tiny CC}}$'s (red symbols) and 
$\sigma_{\text{\tiny R}}$'s (black symbols) of (a) $^{14}$C and (b) $^{16}$C on a $^{12}$C target. 
The red-filled squares are our data at around $45A$ MeV. The red-open squares are the data 
taken from Refs.~\cite{YamaguchiPRT107,ChulkovNPA674}. The black-open circles are the
$\sigma_{\text{\tiny R}}$ data taken from Refs.~\cite{TakechiPHD,OzawaNPA691}. To calculate the 
$\sigma_{\text{\tiny R}}$($E$) and $\sigma_{\text{\tiny CC}}$($E$), and to extract the proton and neutron rms radii,
we assumed HO-type proton-density distributions for the protons and neutrons in $^{14}$C.
We used the $\sigma_{\text{\tiny R}}$ data at $950A$ MeV~\cite{OzawaNPA691} and our $\sigma_{\text{\tiny CC}}$
to determine the HO width parameters for the proton- and neutron-density distributions. We have
avoided using the other $\sigma_{\text{\tiny CC}}$ data shown in Fig.~\ref{fig:cccsc}(a) because we found
systematic deviations from our data for all $^{12-16}$C isotopes. We note that the
$\sigma_{\text{\tiny CC}}$ at around $930A$ MeV from Ref.~\cite{ChulkovNPA674} deviates as much as 7$\%$
from the datum at around $950A$ MeV from Ref.~\cite{WebberPRC41}, which we have used
together with the $\sigma_{\text{\tiny R}}$ at $950A$ MeV to determine the global parameters.
In addition, we have also confirmed that our calculations can reproduce the recent $\sigma_{\text{\tiny CC}}$
data for all carbon isotopes at around $900A$ MeV~\cite{KanungoPRL}. The $\sigma_{\text{\tiny CC}}$($E$) and
$\sigma_{\text{\tiny R}}$($E$) thus calculated are shown by the red-dashed and the black lines in
Fig.~\ref{fig:cccsc}(a).\\
%
% Figure 4
%
\indent For the $^{16}$C isotope, a $^{14}$C-core-plus-two-neutron type nucleon density distribution 
has been suggested. Assuming such density distribution, T.~Zheng {\it et al.} deduced a 
nucleon-density distribution with a relatively long tail~\cite{ZhengNPA709}. Here, as a first trial, 
we assumed the HO-type density distributions similar to $^{14}$C. The proton- and neutron-density 
distributions required to reproduce the experimental $\sigma_{\text{\tiny R}}$ at around $950A$ MeV and our 
$\sigma_{\text{\tiny CC}}$ are shown by the red-dashed and black-solid lines respectively. Obviously, the 
calculated $\sigma_{\text{\tiny R}}$($E$) underestimates the two experimental $\sigma_{\text{\tiny R}}$'s at energies 
below $100A$ MeV. Such deviation is well known and has been observed in the reactions of $^{11}$Be 
on $^{12}$C and $^{27}$Al at $33A$-MeV incident energy~\cite{FukudaPLB268}. To reproduce the 
experimental $\sigma_{\text{\tiny R}}$ at low energy, we considered the HO core plus WS-type-two-neutron tail (core+2n) density 
distribution. The parameters for the core+2n neutron-density distribution 
were determined so as to reproduce the experimental $\sigma_{\text{\tiny R}}$'s at $39A$~\cite{FangPRC61} and 
$950A$~\cite{OzawaNPA691} MeV. The core+2n density distribution thus deduced is shown by 
the black-dotted  line in the inset of Fig.~\ref{fig:cccsc}(b). The calculation 
 also reproduces the experimental data at $83A$ MeV~\cite{ZhengNPA709}
very well. 
We note that similar neutron-density distribution, i.e. HO core plus WS-type-one-neutron tail (core+n), is also required to reproduce the sole $\sigma_{\text{\tiny R}}$ datum for $^{15}$C at around $20A$ MeV~\cite{FangPRC61}. However, the experimental uncertainty is too large to allow any definite conclusion. 
Hence, the present results confirm the importance of the $\sigma_{\text{\tiny R}}$ (and perhaps 
the $\sigma_{\text{\tiny CC}}$) measurements at incident energies below $100A$ MeV.\\
\indent The proton rms radii for $^{12-16}$C thus extracted from our measured $\sigma_{\text{\tiny CC}}$'s
are also shown in Table~\ref{tab:cccs}. For comparison, the experimental proton rms radii for
$^{12-14}$C from the electron-scattering data~\cite{AngeliADNDT99} and $^{15-16}$C from
Ref.~\cite{YamaguchiPRT107} are also shown. The uncertainties of proton rms radii shown in Table~\ref{tab:cccs} are from the uncertainties of $\sigma_{\text{\tiny CC}}$'s, and do not include those of $\beta_{\rm pn}$, which are mainly from $\sigma_{\text{\tiny R}}$'s. In this energy region, the uncertainty of $\sigma_{\text{\tiny R}}$ is almost equivalent to that of our $\sigma_{\text{\tiny CC}}$. Including these uncertainties results in an additional uncertainty factor of about $\sqrt{2}$ in proton rms radii, which will not affect our conclusion. It is important to note that our results for $^{12-14}$C
are in good agreement with, within one standard deviation from, the electron-scattering data.
These agreements provide further justification for the adoption of our experimental
$\sigma_{\text{\tiny CC}}$'s in determining the density distributions. The general consistencies between the
experimental ($\sigma_{\text{\tiny R}}$ and $\sigma_{\text{\tiny CC}}$) cross sections and our Glauber-model
calculations  with global parameters demonstrate the validity
and versatility of the model for various isotopes over a
wide range of incident energies.
\section{Summary}
In summary, we have measured the $\sigma_{\text{\tiny CC}}$'s of $^{12-16}$C on a carbon target using the
transmission method at around 45$A$ MeV incident-beam energies at the RCNP EN course, Osaka
University. To analyze the low-energy data, we have developed a finite-range Glauber model with a global parameter set
within the optical-limit approximation, which is applicable to incident energies below $100A$ MeV.
Our calculations show excellent agreement with the experimental $\sigma_{\text{\tiny R}}$'s for reactions of
$^{12}$C on a $^{9}$Be and $^{27}$Al targets. Performing the Glauber-model analysis on the experimental
$\sigma_{\text{\tiny R}}$ and $\sigma_{\text{\tiny CC}}$, we show the sensitivity of the low-energy
$\sigma_{\text{\tiny R}}$  to the tail-density distribution of ``neutron-halo'' or
``neutron-skin'' nuclei. The results confirm the importance of the $\sigma_{\text{\tiny R}}$ (and
perhaps $\sigma_{\text{\tiny CC}}$) measurements at incident energies below $100A$ MeV. We also extracted
the proton rms radii for $^{12-16}$C using our measured $\sigma_{\text{\tiny CC}}$'s and the existing
$\sigma_{\text{\tiny R}}$ data. The results for $^{12-14}$C are in good agreement with the values from
the electron scatterings. These consistencies, together with the capability of our calculations to
reproduce most of the experimental $\sigma_{\text{\tiny R}}$ and $\sigma_{\text{\tiny CC}}$ data for several isotopes
and over a wide range of incident energies, demonstrate the usefulness of the
$\sigma_{\text{\tiny CC}}$ measurement and our Glauber model.
\section{Acknolwedgements}
We thank the RCNP Ring Cyclotron staff for delivering $^{22}$Ne beam stably throughout the experiment,
and Prof. K.~Kimura for his precious advice during the preparation of the MUSIC detector.
H.J.O. and I.T. would like to acknowledge the support of Prof. A.~Tohsaki (Suzuki) and his 
spouse. D.T.T. and T.T.N. appreciate the support of RCNP Visiting Young Scientist Support Program.
The support of the Vietnam government under the Program of Development in Physics by 2020, 
the PR China government and Beihang University under 
the Thousand Talent Program, and the supports from the Nishimura and Hirose 
International Scholarship Foundations are gratefully acknowledged.
This work was supported in part by JSPS-VAST Bilateral Joint Research Project and Grand-in-Aid for Scientific Research Nos. 20244030, 20740163,  23224008 from Monbukagakusho,
Japan.
% Create the reference section using BibTeX:
%\bibliography{ref.bib}

\end{document}